\documentclass[preprint,aps,prd,amsmath,amssymb]{revtex4}
\pdfoutput=1
\usepackage{graphicx}
\usepackage{color}
\newcommand{\beq}{\begin{eqnarray}}
\newcommand{\eeq}{\end{eqnarray}}

\newcommand{\bmp}{\noindent\begin{minipage}{16cm}}
\newcommand{\emp}{\end{minipage}\vskip 7mm} 

\usepackage{dcolumn}
\usepackage{bm}
\usepackage{bbm}
\usepackage{subfigure}
\usepackage{pxfonts}

\usepackage{epsfig}

\def\lsim{\mathrel{\rlap{\lower4pt\hbox{\hskip1pt$\sim$}}
    \raise1pt\hbox{$<$}}}                
\def\gsim{\mathrel{\rlap{\lower4pt\hbox{\hskip1pt$\sim$}}
    \raise1pt\hbox{$>$}}}                

\newcommand{\drawsquare}[2]{\hbox{%
\rule{#2pt}{#1pt}\hskip-#2pt
\rule{#1pt}{#2pt}\hskip-#1pt
\rule[#1pt]{#1pt}{#2pt}}\rule[#1pt]{#2pt}{#2pt}\hskip-#2pt
\rule{#2pt}{#1pt}}

\newcommand{\Yfund}{\raisebox{-.5pt}{\drawsquare{6.5}{0.4}}}
\newcommand{\Ysymm}{\raisebox{-.5pt}{\drawsquare{6.5}{0.4}}\hskip-0.4pt%
        \raisebox{-.5pt}{\drawsquare{6.5}{0.4}}}
\newcommand{\Ythrees}{\raisebox{-.5pt}{\drawsquare{6.5}{0.4}}\hskip-0.4pt%
          \raisebox{-.5pt}{\drawsquare{6.5}{0.4}}\hskip-0.4pt%
          \raisebox{-.5pt}{\drawsquare{6.5}{0.4}}}
\newcommand{\Yasymm}{\raisebox{-3.5pt}{\drawsquare{6.5}{0.4}}\hskip-6.9pt%
        \raisebox{3pt}{\drawsquare{6.5}{0.4}}}
\newcommand{\Ythreea}{\raisebox{-3.5pt}{\drawsquare{6.5}{0.4}}\hskip-6.9pt%
        \raisebox{3pt}{\drawsquare{6.5}{0.4}}\hskip-6.9pt
        \raisebox{9.5pt}{\drawsquare{6.5}{0.4}}}

\newcommand{\Yadjoint}{\raisebox{-3.5pt}{\drawsquare{6.5}{0.4}}\hskip-6.9pt%
        \raisebox{3pt}{\drawsquare{6.5}{0.4}}\hskip-0.4pt
        \raisebox{3pt}{\drawsquare{6.5}{0.4}}}
\begin{document}
\title{\Large  \color{red} Higher Representations Duals}
\author{Francesco {\sc Sannino}$^{\color{blue}{\varheartsuit}}$}
\email{sannino@cp3.sdu.dk} 
\affiliation{
$^{\color{blue}{\varheartsuit}}${ CP}$^{ \bf 3}${-Origins}, 
Campusvej 55, DK-5230 Odense M, Denmark.\footnote{{ C}entre of Excellence for { P}article { P}hysics { P}henomenology devoted to the understanding of the {Origins} of Mass in the universe. }}
\begin{flushright}
{\it CP$^3$- Orgins: 2009-13}
\end{flushright}
\begin{abstract}
We uncover novel solutions of the 't Hooft anomaly matching conditions for scalarless gauge theories with matter transforming according to higher dimensional representations of the underlying gauge group. We argue that, if the duals exist, they are gauge theories with fermions transforming according to the defining representation of the dual gauge group. The resulting conformal windows match the one stemming from the all-orders beta function results when taking the anomalous dimension of the fermion mass to be unity which are also very close to the ones obtained using the Schwinger-Dyson approximation. We use the solutions to gain useful insight on the conformal window of the associated electric theory. A consistent picture emerges corroborating previous results obtained via different analytic methods and in agreement with first principle lattice explorations. \end{abstract}

\maketitle

\section {Introduction}

One of the most fascinating possibilities is that generic asymptotically free gauge theories have magnetic duals. In fact, in the late nineties, in a series of  ground breaking papers Seiberg  \cite{Seiberg:1994bz,Seiberg:1994pq} provided strong support for the existence of a consistent picture of such a duality within a supersymmetric framework. Supersymmetry is, however, quite special and the existence of such a duality does not automatically imply the existence of nonsupersymmetric duals. One of the most relevant  results put forward by Seiberg  has been the identification of the boundary of the conformal window for supersymmetric QCD as function of the number of flavors and colors. 
The dual theories proposed by Seiberg pass a set of mathematical consistency relations known as 't Hooft anomaly conditions \cite{Hooft}.  Another important tool has been the knowledge of the all-orders supersymmetric beta function \cite{Novikov:1983uc,Shifman:1986zi,Jones:1983ip}

Recently we provided several analytic predictions for the conformal window of nonsupersymmetric gauge theories using different approaches \cite{Sannino:2004qp,Dietrich:2006cm,Ryttov:2007cx}. 
We also initiated in \cite{Sannino:2009qc}  the exploration of the possible existence of truly QCD nonsupersymmetric gauge dual providing a consistent picture of the phase diagram as function of number of colors and flavors. 

Arguably the existence of a possible dual of a generic nonsupersymmetric asymptotically free gauge theory able to reproduce its infrared dynamics must match the 't Hooft anomaly conditions \cite{Hooft}. We have exhibited several solutions of these conditions for QCD in \cite{Sannino:2009qc}. An earlier exploration already appeared in the literature \cite{Terning:1997xy}. The novelty with respect to these earlier results were: i) The request that the gauge singlet operators associated to the magnetic baryons should be interpreted as bound states of ordinary baryons \cite{Sannino:2009qc}; ii) The fact that the asymptotically free condition for the dual theory matches the lower bound on the conformal window obtained using the all-orders beta function  \cite{Ryttov:2007cx}. These extra constraints help restricting further the number of possible gauge duals without diminishing the exactness of the associate solutions with respect to the 't Hooft anomaly conditions. 

In this paper we analyze theories with fermions transforming according to higher dimensional representations. Some of these theories have been used to construct sensible extensions of the standard model of particle interactions of technicolor type passing precision data and known as Minimal Walking Technicolor models \cite{Sannino:2004qp,Dietrich:2006cm}. Other interesting studies of technicolor dynamics making use of higher dimensional representations appeared in \cite{Christensen:2005cb}. These are the only known extension of technicolor type possessing the smallest naive $S$ parameter while being able to display simultaneously (near) conformal behavior.  One can also construct explicit examples of extended technicolor interactions \cite{Eichten:1979ah} for these models \cite{Evans:2005pu}.

We will exhibit here novel solutions to the 't Hooft anomaly conditions for theories with higher dimensional representations. The resuling {\it magnetic} dual allows to predict the critical number of flavors above which the asymptotically free theory, in the electric variables, enters the conformal regime as predicted using the all-orders conjectured beta function \cite{Ryttov:2007cx}. 

 Several analytic predictions for the lower end of the conformal window for nonsupersymmetric gauge theories with matter transforming according to various $SU$,  $SO$ and $Sp$ representations have been made \cite{Sannino:2004qp,Dietrich:2006cm,Ryttov:2007cx,Sannino:2009aw,Ryttov:2009yw}.  Here we show that  using {\it exact} anomaly matching conditions together with the new constraints coming from operator matching introduced in \cite{Sannino:2009qc} we arrive at a bound on the conformal windows in agreement with earlier analysis. In particular we find our results to agree with the ones obtained using the all-orders beta function conjecture when assuming the maximum anomalous dimension of the fermion mass to be unity. 
 An interesting result, stemming from the restriction that the magnetic baryons have baryonic charges which are multiple of the ordinary baryons, is that the duals prefer to feature magnetic fermions transforming according to the fundamental representation of the dual gauge group.
 
 We begin our analysis by investigating an $SU(3)$ gauge theory with fermionic matter in the two-index symmetric representation of the gauge group. The interest for this theory resides in the fact that it has been used to construct models of dynamical electroweak symmetry breaking. It is also being studied via first principle lattice simulations. We first investigate the possible duals with fermions in the defining representation of the dual gauge group and then analyze the case in which the magnetic fermions transform according to the same higher dimensional representation. We find that operator matching privileges the former duals. Here the various consistency conditions indicate the lower bound of the conformal window must start already for two Dirac flavors. The results are in surprising agreement with the ones derived using the all-orders beta function with an anomalous dimension of the fermion mass near unity, and as well as the ones  first predicted in  \cite{Sannino:2004qp} and obtained via the Schwinger-Dyson (SD) approximation \cite{Appelquist:1988yc,Cohen:1988sq,Appelquist:1996dq,Miransky:1996pd}. 
 
 We then investigate $SU(N)$ gauge theories with adjoint Weyl matter. Here we find that an interesting dual is an $SO(N_f -1)$ gauge theory with $N_f$ Weyl fermions in the vector representation. In this case the number of Weyl fermoins above which we expect conformality to set in is $N_f =4$, corresponding to exactly $2$ Dirac flavors. This result is in agreement with analytical \cite{Sannino:2004qp,Dietrich:2006cm,Ryttov:2007cx,Sannino:2009aw, Poppitz:2009uq} and first principle lattice simulations  \cite{Catterall:2007yx,Catterall:2008qk,DelDebbio:2008wb,Shamir:2008pb,Hietanen:2008mr,DelDebbio:2008tv,Hietanen:2009az,DelDebbio:2009fd}. Our results can be further tested via first principle lattice simulations and strongly reduce the number of possible gauge theories one can use to construct models of nature given that the underlying dynamics of superficially distinct models may be related, in the infrared,  by a gauge duality. Currently the conformal windows of exactly the theories investigated here, as well as the one investigated in \cite{Sannino:2009qc}, are object of intense analytic  \cite{Sannino:2004qp,Dietrich:2006cm,Ryttov:2007cx,Sannino:2009aw, Poppitz:2009uq, Armoni:2009jn, Golkar:2009aq,Dietrich:2009ns} and numerically oriented  \cite{Catterall:2007yx,Catterall:2008qk,DelDebbio:2008wb,Shamir:2008pb,Hietanen:2008mr,DelDebbio:2008tv,Hietanen:2009az,DelDebbio:2009fd, Fodor:2009ar,Fodor:2009wk,  Fodor:2009nh,Fodor:2008hn, Sinclair:2009ec, Appelquist:2007hu,Deuzeman:2008sc,DeGrand:2008kx,Appelquist:2009ty,Deuzeman:2009mh,DeGrand:2009et,Hasenfratz:2009ea} investigations.

\section{SU(3) Gauge Theory with $N_f$ Dirac Flavors in the 2-index symmetric representation} 
The underlying gauge group is $SU(3)$ while the
quantum flavor group is
\begin{equation}
SU_L(N_f) \times SU_R(N_f) \times U_V(1) \ ,
\end{equation}
and the classical $U_A(1)$ symmetry is destroyed at the quantum
level by the Adler-Bell-Jackiw anomaly. We indicate with
$Q_{\alpha;\{c_1,c_2\}}^i$ the two component left spinor where $\alpha=1,2$
is the spin index, $c_1, c_2=1,...,3$ is the color index while
$i=1,...,N_f$ represents the flavor. $\widetilde{Q}^{\alpha ;\{c_1,c_2\}}_i$
is the two component conjugated right spinor. We summarize the
transformation properties in the following table.
\begin{table}[h]
\[ \begin{array}{|c| c | c c c|} \hline
{\rm Fields} &  \left[ SU(3) \right] & SU_L(N_f) &SU_R(N_f) & U_V(1) \\ \hline \hline
Q &\Ysymm &{\Yfund }&1&~~1  \\
\widetilde{Q} & \overline{\Ysymm}&1 &  \overline{\Yfund}& -1   \\
G_{\mu}&{\rm Adj}   &1&1  &~~1\\
 \hline \end{array} 
\]
\caption{Field content of an SU(3) gauge theory with quantum global symmetry $SU_L(N_f)\times SU_R(N_f) \times U_V(1)$. }
\end{table}

The  global anomalies are associated to the triangle diagrams featuring at the vertices three $SU(N_f)$ generators (either all right or all left), or two 
$SU(N_f)$ generators (all right or all left) and one $U_V(1)$ charge. We indicate these anomalies for short with:
\begin{equation}
SU_{L/R}(N_f)^3 \ ,  \qquad  SU_{L/R}(N_f)^2\,\, U_V(1) \ .
\end{equation}
For a vector like theory there are no further global anomalies. The
cubic anomaly factor, for fermions in fundamental representations,
is $1$ for $Q$ and $-1$ for $\tilde{Q}$ while the quadratic anomaly
factor is $1$ for both leading to
\begin{equation}
SU_{L/R}(N_f)^3 \propto \pm 6 \ , \quad SU_{L/R}(N_f)^2 U_V(1)
\propto \pm 6 \ .
\end{equation}
 
 \begin{figure}[h!]
\includegraphics[width=12cm]{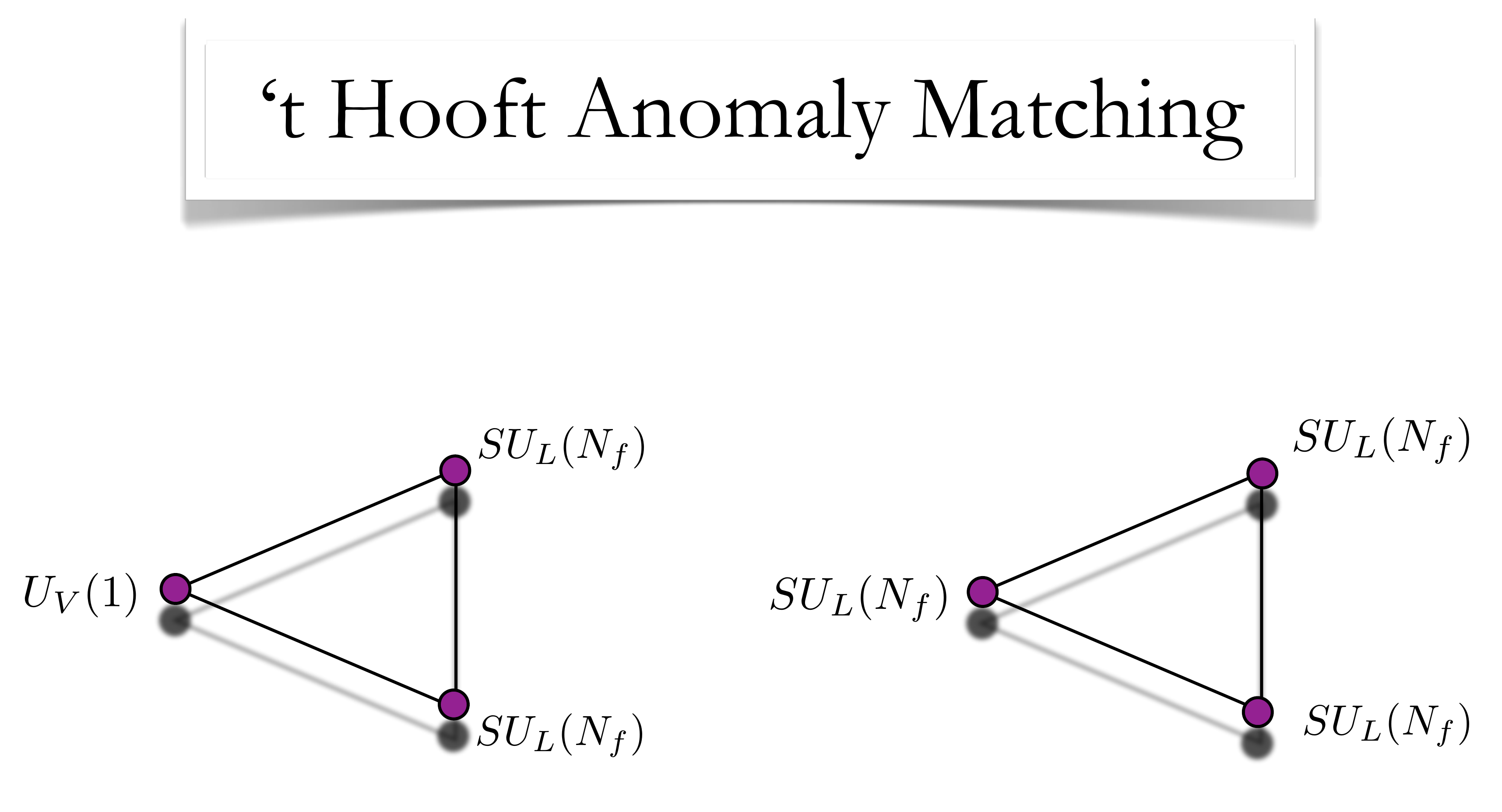}
\caption{The 't Hooft anomaly matching conditions are related to the saturation of the global anomalies stemming out of the one-loop triangle diagrams represented, for the theory of interest, here. According to 't Hooft both theories, i.e. the electric and the magnetic ones, should yield the same global anomalies.}
\label{GA}
\end{figure}
\subsection{Conformal window from the all orders beta function}
Recently we have conjectured an all-orders beta function which allows for a bound of the conformal window \cite{Ryttov:2007cx} of gauge theories for any matter representation.  Other approaches yield compatible results. 

In this paper we show that thanks to the identification of new possible gauge duals we lend further strong support in favor of this conjecture. Consider an $SU(N)$  gauge group with $N_f$ Dirac flavors belonging to the representation $r$ of the gauge group. 
The conjectured beta function \cite{Ryttov:2007cx} is given in terms of the anomalous dimension of the fermion mass $\gamma=-{d\ln m}/{d\ln \mu}$ where $m$ is the renormalized mass. At the zero of the all-orders beta function one has
\begin{eqnarray}
 \frac{2}{11}T(r)N_f(r) \left( 2+ \gamma \right) = C_2(G) \ ,
\end{eqnarray}
The generators $T_r^a,\, a=1\ldots N^2-1$ of the gauge group in the
representation $r$ are normalized according to
$\text{Tr}\left[T_r^aT_r^b \right] = T(r) \delta^{ab}$ while the
quadratic Casimir $C_2(r)$ is given by $T_r^aT_r^a = C_2(r)I$. The
trace normalization factor $T(r)$ and the quadratic Casimir are
connected via $C_2(r) d(r) = T(r) d(G)$ where $d(r)$ is the
dimension of the representation $r$. The adjoint
representation is denoted by $G$. 
Hence, specifying the value of the anomalous dimensions at the IRFP yields the last constraint needed to construct the conformal window. Requiring the absence of negative norm states  at the conformal point requires  $\gamma < 2$ resulting in the {\it maximum} possible extension of the conformal window bounded from below by:
\begin{equation}
N_f(r)^{\rm BF} \geq \frac{11}{8} \frac{C_2(G)}{T(r)}  \qquad { \gamma =2}\ .
\end{equation}
The actual size of the conformal window can, however, be smaller than the one determined above without affecting the validity of the beta function. It may happen, in fact, that chiral symmetry breaking is triggered for a value of the anomalous dimension less than two. If this occurs the conformal window shrinks. The ladder approximation approach \cite{Appelquist:1988yc,{Cohen:1988sq},Appelquist:1996dq,Miransky:1996pd},  for example, predicts that chiral symmetry breaking occurs when the anomalous dimension is larger than one. Remarkably the all-orders beta function encompass this possibility as well \cite{Ryttov:2007cx}. In fact, it is much more practical to quote the value predicted using the beta function by imposing $\gamma =1$:
\begin{eqnarray}\label{Gamma-One}
N_f(r)^{\rm BF} \geq  \frac{11}{6}  \frac{C_2(G)}{T({r})} \ ,\qquad {\gamma =1}  \ .
\end{eqnarray}
The result is very close to the one obtained using directly the ladder approximation as shown in \cite{Ryttov:2007cx,Sannino:2009aw}.

 Lattice simulations of the conformal window for various matter representations  \cite{Catterall:2007yx,Catterall:2008qk,Shamir:2008pb,DelDebbio:2008wb,
 Hietanen:2008mr,Appelquist:2007hu,Deuzeman:2008sc,Fodor:2008hn,DelDebbio:2008tv,DeGrand:2008kx,Appelquist:2009ty,Hietanen:2009az,Deuzeman:2009mh,DeGrand:2009et,Hasenfratz:2009ea} are in agreement with the predictions of the conformal window via the all-orders beta function.

\subsection{SU(N) with 2-index symmetric matter conformal window via BF}

Specializing to $SU(N)$ with two-index symmetric representation we find: 
\begin{equation}
N_f(r)^{\rm BF} \geq \frac{11N}{4(N+2)}  \ , \qquad SU(N)~{\rm for~Symmetric~rep.~with}  \quad  { \gamma =2}  .
\end{equation}
which for $N=3$ implies $N_f(r)^{\rm BF} \geq 1.65$.

Assuming, instead, the lower bound to occur for $\gamma =1 $ we discover that:
\begin{equation}
N_f(r)^{\rm BF} \geq \frac{11N}{3(N+2)}  \ , \qquad SU(N)~{\rm for~Symmetric~rep.~with}  \quad  { \gamma =1}  .
\end{equation}
which for $N=3$ implies: $N_f(r)^{\rm BF} \geq 2.2 $.

{}It is desirable to have a novel way to determine the conformal window which makes use of exact matching conditions.  By comparing the various methods one can infer the anomalous dimension to pick as boundary of the window.

\section{Dual of the SU(3) and 2-index Symmetric theory} 

\begin{figure}[h!]
\includegraphics[width=12cm]{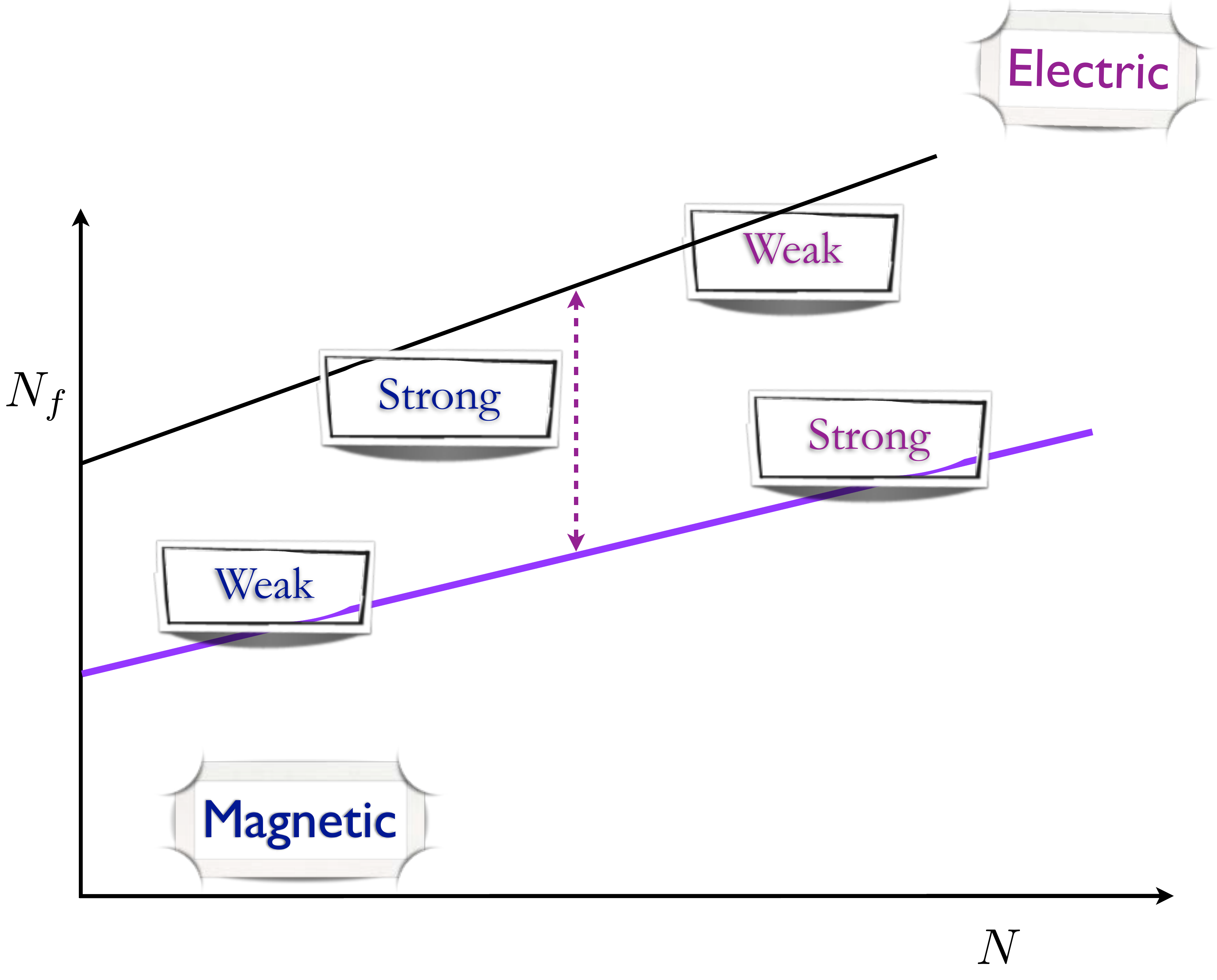}
\caption{Schematic representation of the phase diagram as function of number of flavors and colors. For a given number of colors by increasing the number flavors within the conformal window we move from the lowest line (violet) to the upper (black) one. The upper black line corresponds to the one where one looses asymptotic freedom in the electric variables and the lower line where chiral symmetry breaks and long distance conformality is lost. In the {\it magnetic} variables the situation is reverted and the perturbative line, i.e. the one where one looses asymptotic freedom in the magnetic variables, correspond to the one where chiral symmetry breaks in the electric ones. }
\label{Duality}
\end{figure}

If a magnetic dual does exist one expects it to be weakly coupled near the critical number of flavors below which one breaks  large distance conformality in the electric variables. This idea is depicted in Fig~\ref{Duality}. 

Unfortunately the saturation of the global anomalies is an important tool but is not able to select out a unique solution. We shall see, however, that one class of solutions, when interpreted as containing a possible dual, leads to a prediction of a critical number of flavors corresponding exactly to the one obtained via the conjectured all-orders beta function.

 We seek solutions of the anomaly matching conditions for a gauge theory $SU(X)$ with global symmetry group $SU_L(N_f)\times SU_R(N_f) \times U_V(1)$  featuring 
{\it magnetic} quarks ${q}$ and $\widetilde{q}$ together with $SU(X)$ gauge singlet states identifiable as baryons built out of the {\it electric} quarks $Q$. Since mesons do not affect directly global anomaly matching conditions we could add them to the spectrum of the dual theory.  We study the case in which $X$ is a linear combination of number of flavors and colors of the type $\alpha N_f + N \beta$ with $\alpha$ and $\beta$ integer numbers. In fact, in the following we will consider $N=3$. We will also require that the baryons constructed out of the {\it magnetic} quarks have integer baryonic charges with respect to the original baryon number. In this way they will be interpreted as possible bound states of the original baryons. We will see that this is an important property helping selecting possible duals.

\subsection{Dual Quarks in the Fundamental Representation}

In this initial investigation we search for dual electric quarks in the fundamental representation of the gauge group $X$. This choice has the virtue to keep linear in $N_f$ the asymptotic freedom condition we will investigate later. We have searched for the more complicate case of dual fermions in higher dimensional representaitons and will present this possibility in the following  section. 

We add to the {\it magnetic} quarks gauge singlet Weyl fermions which can be identified with massless baryons of the electric theory. The generic dual spectrum is summarized in table \ref{dualgeneric2indices}.
\begin{table}[h]
\[ \begin{array}{|c| c|c c c|c|} \hline
{\rm Fields} &\left[ SU(X) \right] & SU_L(N_f) &SU_R(N_f) & U_V(1)& \# ~{\rm  of~copies} \\ \hline 
\hline 
 q &\Yfund &{\Yfund }&1&~~y &1 \\
\widetilde{q} & \overline{\Yfund}&1 &  \overline{\Yfund}& -y&1   \\
A &1&\Ythreea &1&~~~3& \ell_A \\
S &1&\Ythrees &1&~~~3& \ell_S \\
C &1&\Yadjoint &1&~~~3& \ell_C \\
B_A &1&\Yasymm &\Yfund &~~~3& \ell_{B_A} \\
B_S &1&\Ysymm &\Yfund &~~~3& \ell_{B_S} \\
{D}_A &1&{\Yfund} &{\Yasymm } &~~~3& \ell_{{D}_A} \\
{D}_S & 1&{\Yfund}  &{\Ysymm} &  ~~~3& \ell_{{D}_S} \\
\widetilde{A} &1&1&\overline{\Ythreea} &-3&\ell_{\widetilde{A}}\\
\widetilde{S} &1&1&\overline{\Ythrees} & -3& \ell_{\widetilde{S}} \\
\widetilde{C} &1&1&\overline{\Yadjoint} &-3& \ell_{\widetilde{C}} \\
 \hline \end{array} 
\]
\caption{Massless spectrum of {\it magnetic} quarks and baryons and their  transformation properties under the global symmetry group. The last column represents the multiplicity of each state and each state is a  Weyl fermion.}
\label{dualgeneric2indices}
\end{table}
The wave functions for the gauge singlet fields $A$, $C$ and $S$ are obtained by projecting the flavor indices of the following operator
\begin{eqnarray}
\epsilon^{c_1c_2c_3}\epsilon^{d_1 d_2 d_3} Q_{\{c_1,d_1\}}^{i_1} Q_{\{c_2,d_2\}}^{i_2} Q_{\{c_3,d_3\}}^{i_3}\ ,
\end{eqnarray}
over the three irreducible representations of $SU_L(N_f)$ as indicated in the table \ref{dualgeneric}. These states are all singlets under the $SU_R(N_f)$ flavor group. Similarly one can construct the only right-transforming baryons $\widetilde{A}$, $\widetilde{C}$ and $\widetilde{S}$ via $\widetilde{Q}$. The $B$ states are made by two $Q$ fields and one right field $\overline{\widetilde{Q}}$ while the $D$ fields are made by one $Q$ and two $\overline{\widetilde{Q}}$ fermions. $y$ is the, yet to be determined, baryon charge of the {\it magnetic} quarks while the baryon charge of composite states is fixed in units of the electric quark one. The $\ell$s count the number of times the same baryonic matter representation appears as part of the spectrum of the theory. Invariance under parity and charge conjugation of the underlying theory requires $\ell_{J} = \ell_{\widetilde{J}}$~~ with $J=A,S,...,C$ and $\ell_B = - \ell_D$. 

Having defined the possible massless matter content of the gauge theory dual to the electric theory we compute the $SU_{L}(N_f)^3$ and $SU_{L}(N_f)^2\,\, U_V(1)$ global anomalies in terms of the new fields:   
 \begin{eqnarray}
SU_{L}(N_f)^3 &\propto &  {X } + \frac{(N_f-3)(N_f -6)}{2}\,\ell_A + \frac{(N_f+3)(N_f +6)}{2}\,\ell_S + (N_f^2 - 9)\,\ell_C\nonumber \\ &&   +\,(N_f-4)N_f\,\ell_{B_A} + \,(N_f+4)N_f\,\ell_{B_S}  + \frac{N_f(N_f-1)}{2}\,\ell_{D_A} \nonumber \\ 
&&+\frac{N_f(N_f+1)}{2}\,\ell_{D_S}  = 6 \ ,  \\ 
 & &\\
SU_{L}(N_f)^2\,\, U_V(1) &\propto &  y\, {X} +3 \frac{(N_f-3)(N_f -2)}{2}\,\ell_A + 3\frac{(N_f+3)(N_f +2)}{2}\,\ell_S + 3 (N_f^2 - 3)\,\ell_C \nonumber \\ 
 &&  +\, 3(N_f-2)N_f\,\ell_{B_A} + \,3(N_f+2)N_f\,\ell_{B_S}  + 3\frac{N_f(N_f-1)}{2}\,\ell_{D_A}\nonumber \\ &&+3\frac{N_f(N_f+1)}{2}\,\ell_{D_S} =6  \ .   \end{eqnarray}
  The left-hand expressions are identical to the ones of QCD while the right-hand side provides the corresponding value of the anomaly for the electric theory  with two-index symmetric matter.  
  
\begin{figure}[h!]
\includegraphics[width=12cm]{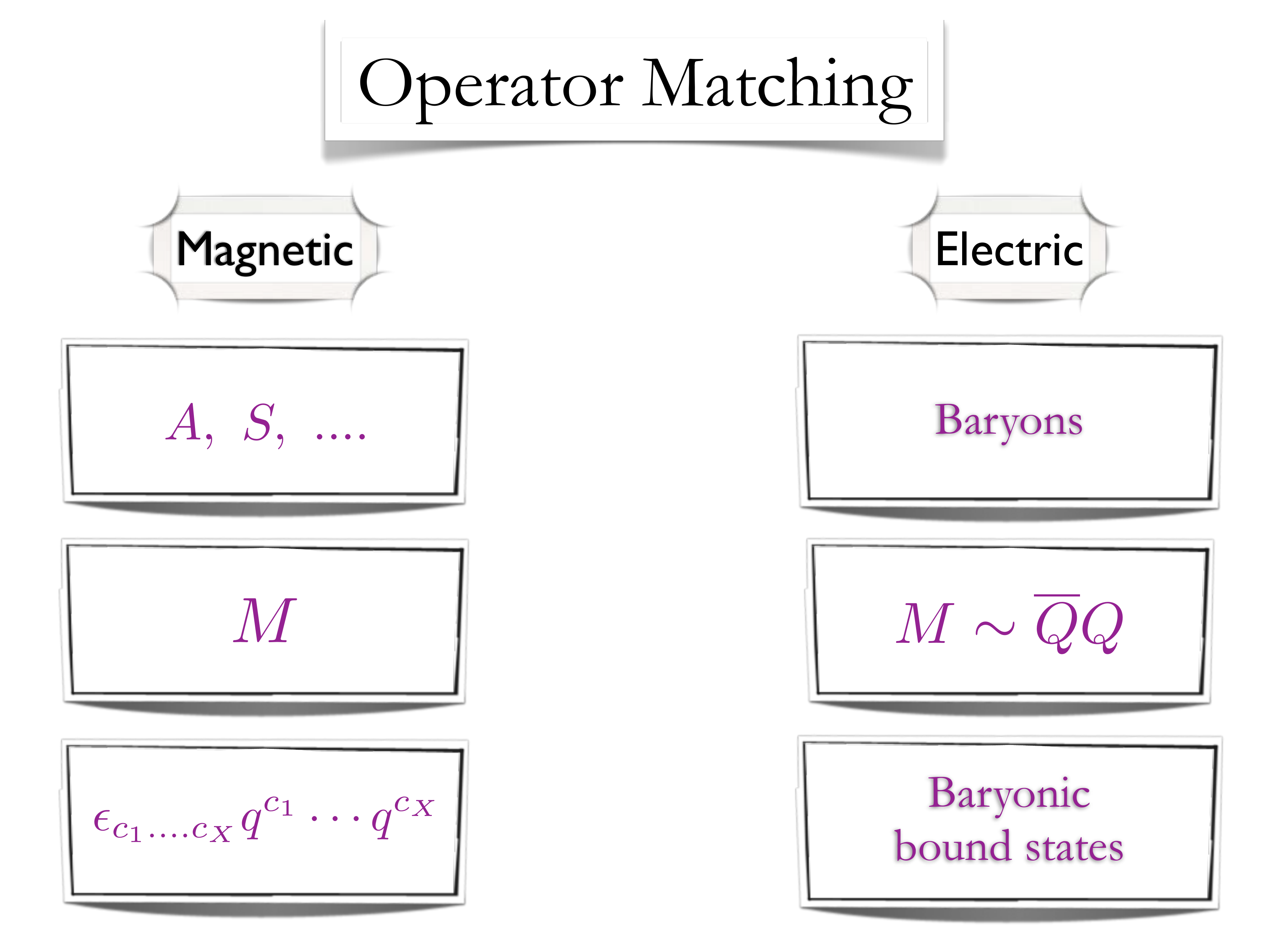}
\caption{We propose the above correspondence between the gauge singlet operators of the magnetic theory and the electric ones. The novelty introduced in \cite{Sannino:2009qc} with respect to any of the earlier approaches is the identification of the {\it magnetic} baryons, i.e. the ones constructed via the magnetic quarks, with bound states of baryons in the electric variables.}
\label{OpM}
\end{figure}

  In Seiberg's analysis it was also possible to match some of the operators of the magnetic theory with the ones of the electric theory. The situation for nonsupersymmetric theories is, in principle, more involved although it is clear that certain magnetic operators match exactly the respective ones in the electric variables. These are the meson $M$ and the massless baryons, $A$, $\widetilde{A}$, ...., $S$ shown in Table \ref{dualgeneric}.  Besides these obvious identifications we also required  \cite{Sannino:2009qc}  that the baryonic type operators constructed via the magnetic dual quarks should have baryonic charges multiple of the ordinary baryons ones. We proposed to identify them, in the electric variables, with bound states of ordinary baryons. We summarize the proposed operator matching constraints in Fig.~\ref{OpM}.
  
  In the case of $N_f =2 $ the cubic anomaly vanishes identically and should not be considered. {}For three colors the electric theory looses asymptotic freedom for $3.3$ flavors and hence there is only one value of $N_f$, i.e. $N_f=3$ for which both anomalies are relevant. 
 
We have found different solutions to the anomaly matching conditions which we will present the:
\subsubsection{First solution: $SU(2N_F -3)$ dual gauge group   }
The solutions correspond, for the case $N=3$ to the following value assumed by the indices and $y$ baryonic charge: 
  \begin{eqnarray}
 X &= &2N_f- 3\ , \quad \ell_{A}= 0  \ ,  \quad \ell_{D_A} =k_1 = -\ell_{B_A}  \ , \nonumber \\
 && \nonumber \\ 
   \ell_{S}&=& -1 + 2k_1 + 5k_2 \ , \quad  \ell_{D_S} =  -2 + 4 k_1 + 9 k_2 = - \ell_{B_S} \ , \nonumber \\ 
   && \nonumber \\
  \ell_C  & = & 0 \ , \qquad   y = 3\, \frac{a + bN_f + c N_f ^2}{ 4 N_f - 6}  \ ,  \nonumber 
   \end{eqnarray}
 with $k_1$ and $k_2$ integer numbers and  \begin{equation}
 a =  10-12 k_1 - 30 k_2, \quad b= 2k_2  -  k_1  - 1  \quad   c= 3k_1 + 4 k_2 -1 \ .
 \end{equation}
We have asked that both anomaly matching conditions are satisfied for $N_f =3$  and that the solutions satisfy also the quadratic one for $N_f=2$. 

Of course $X$ must assume a value strictly larger than one otherwise it describes an abelian gauge theory. This provides the first nontrivial bound on the number of flavors: 
 \begin{equation}
 N_f > \frac{3+ 1}{2} = 2  \ . \end{equation} 
This value is remarkably  consistent with the maximum extension predicted using the truncated SD equation and the all-orders beta function for a value of the anomalous dimension equal to one. 

Asymptotic freedom of the newly found theory is dictated by the coefficient of the one-loop beta function :
 \begin{equation}
 \beta_0 = \frac{11}{3} (2N_f- 3)  - \frac{2}{3}N_f\ . 
 \end{equation}
 To this order in perturbation theory the gauge singlet states do not affect the {magnetic} quark sector and we can hence determine  the number of flavors obtained by requiring the dual theory to be asymptotic free. i.e.: 
\begin{equation}
N_f \geq \frac{33}{20} = 1.65 \ , \qquad\qquad\qquad {\rm Dual~Asymptotic~Freedom}  \qquad {\rm I}\ . 
\end{equation}
This value {\it coincides} with the one predicted by means of the all-orders conjectured beta function for the lowest bound of the conformal window, in the {\it electric} variables, when taking the anomalous dimension of the mass to be $\gamma =2 $. We recall that for any number of colors $N$ the all orders beta function requires the critical number of flavors to be larger than: 
\begin{equation}
N_f^{BF}|_{\gamma = 2} = \frac{11N}{4(N+2)}  \ . 
\end{equation}
{}For N=3 the two expressions yield $1.65$. Actually given that $X$ must be larger than one this solution requires $N_f > 2$ rather than $1.65$. This last feature was also observed for the QCD dual case. We simply consider this as a signal that we cannot arrive at the maximum value of $\gamma$, nevertheless we can still arrive at a value for the anomalous dimension larger than one according to this solution. If one requires an even more stringent constraint $X \geq 2$ we then find $N_f>2.5$ which is very close to the result obtained setting $\gamma=1$ in the all-orders beta function.

The baryon charge of the magnetic baryons is: 
\begin{equation}
B[q^X] = X \times y = \frac{3}{2} (a+bN_f + c N_f^2)   = {\rm Operator~matching} = 3\, n  \ ,
\end{equation}
with $n$ an integer requiring  $a+bN_f + c N_f^2$ to be an even number. This extra constraints is easily satisfied by choosing, for example, $k_1 =1$ and $k_2 = 0$ yielding $B[q^X] = 3\, (N_f^2 - N_f  -1)$. {}Intriguingly for $N_f=2$ one recovers the standard baryonic charge. 


%



%
%

\subsubsection{Second solution: SU(7Nf - 15)}

The solutions correspond to the following value assumed by the indices and $y$ baryonic charge: 
  \begin{eqnarray}
 X &= &7N_f- 15\ , \quad \ell_{A}=0 \ ,  \quad \ell_{D_A} = k_1  = -\ell_{B_A}  \ , \nonumber \\
 && \nonumber \\ 
   \ell_{S}&=& 2k_1 + 5k_2 \ , \quad  \ell_{D_S} =    4 k_1 + 9 k_2  = - \ell_{B_S} \ , \nonumber \\ 
   && \nonumber \\
   \ell_C  & = & 0 \ , \qquad 
  y = 3 \frac{a+ bN_f +c N_f^2}{14 N_f  - 30}  \ .\nonumber \\
   && \ \ 
 \end{eqnarray}
 with $k_1$ and $k_2$ integer numbers and 
 \begin{equation}
a = 4- 12 k_1 - 30 k_2 \ , \quad b= 2 k_2  - k_1 \ , \quad c= 3 k_1 +4 k_2  \ .
\end{equation}
The baryon charge of the magnetic baryons is: 
\begin{equation}
B[q^X] = X \times y = \frac{3}{2} (a+bN_f + c N_f^2)   = {\rm Operator~matching} = 3\, n  \ ,
\end{equation}
with $n$ an integer requiring  $a+bN_f + c N_f^2$ to be an even number. This extra constraints is also easily satisfied by choosing, for example, $k_1 =k_2 =0$  yielding $B[q^X] = 6$ for any $N_f$ corresponding to a di-baryon charge. One can also consider the case $k_2 = 1$ and $k_1=0$. 

The condition $X>1$ yields: 
 \begin{equation}
 N_f > \frac{16}{7} \simeq 2.29  \ . \end{equation} 
This value is also remarkably  consistent with the maximum extension predicted using the truncated SD equation and the all-orders beta function for a value of the anomalous dimension equal to one. 

Asymptotic freedom of the newly found theory is dictated by the coefficient of the one-loop beta function :
 \begin{equation}
 \beta_0 = \frac{11}{3} (7N_f- 15)  - \frac{2}{3}N_f\ , 
 \end{equation}
 yielding
 \begin{equation}
N_f \geq \frac{11}{5} = 2.2\ , \qquad\qquad\qquad {\rm Dual~Asymptotic~Freedom} \qquad {\rm II}\ . 
\end{equation}
This value {\it coincides} with the one predicted by means of the all-orders conjectured beta function for the lowest bound of the conformal window, in the {\it electric} variables, when taking the anomalous dimension of the mass to be $\gamma =1 $. We recall that for any number of colors $N$ the all orders beta function requires the critical number of flavors to be larger than: 
\begin{equation}
N_f^{BF}|_{\gamma = 1} = \frac{11N}{3(N+2)}  \ . 
\end{equation}
{}For N=3 the two expressions yield $2.2$.  This value is even closer to the one obtained imposing the condition $X>1$ which is  circa $2.29$. 

Interestingly the two class of solutions suggest that the electric theory is not conformal but walking. We observe that the predictions from the dual theory does {\it not} depend on the all orders beta function. However it is remarkable that the predictions are very close to the ones predicted using the beta function ansatz.  

An interesting property of this solution is that one can saturate the anomaly matching conditions directly via the presumed magnetic quarks. It is, in fact, sufficient to set $k_1 = k_2 =0$ to see this. If we apply the all-order beta function we can investigate when chiral symmetry is restored. Setting $\gamma=1$ for the dual theory one finds that $N_f$ should be less than or equal to about $2.29$ which is lower than the value for which the electric theory looses asymptotic freedom. This seems to indicate  that more matter is needed and the solution $k_1=k_2=0$ is not an exact dual according to the all orders beta function. However one can investigate the nonperturbative dynamics of this theory via first principle lattice simulations and test the duality independently.   

\subsection{Third solution: $SU(\alpha N_f (N+2)  - \beta N + \delta)$ and two-index symmetric magnetic quarks. }

We have investigated also the case in which the magnetic quarks are in the same two-index representation of the gauge group $X = \alpha N_f (N+2)  - \beta N + \delta$. In this case the first coefficient of both anomalies must be modified according to $X \rightarrow X(X+1)/2$ to take into account of the change of the representation of the dual quarks.  We have found several solutions for different integer values of the coefficients $\alpha$, $\beta$ and $\delta$. We present two examples here for $N=3$: 

\subsubsection{$\alpha =2$, $\beta=5$ and $\delta=0$}
 \begin{eqnarray}
 X &= &10 N_f- 15 \ , \quad \ell_{A}=0 \ ,  \quad \ell_{D_A} = k_1  = -\ell_{B_A}  \ , \nonumber \\
 && \nonumber \\ 
   \ell_{S}&=&  -2 + 2k_1 + 5k_2 \ , \quad  \ell_{D_S} =  4 +  4 k_1 + 9 k_2  = - \ell_{B_S} \ , \nonumber \\ 
   && \nonumber \\
   \ell_C  & = & 0 \ , \qquad 
  y = 3 \frac{a+ bN_f +c N_f^2}{10(2 N_f  - 3)(5N_f - 7)}  \ .\nonumber \\
   && \ \ 
 \end{eqnarray}
 with $k_1$ and $k_2$ integer numbers and  
 \begin{equation}
a = 16- 12 k_1 - 30 k_2 \ , \quad b= 22 + 2 k_2  - k_1 \ , \quad c= 6 + 3 k_1 +4 k_2  \ .
\end{equation}

The one-loop coefficient of the beta function is: 
 \begin{eqnarray}
  \beta_0 = \frac{11}{3}X   - \frac{2}{3}N_f  (X +2) \ . 
 \end{eqnarray}
 Asymptotic freedom requires the previous coefficient to be positive which means: 
 \begin{eqnarray}
 1.58 \leq  N_f   \leq 5.22, \quad  {\rm Dual~asymptotic~freedom~condition}\ .
 \end{eqnarray}
 The lower bound is now close to the value of the critical number of flavors corresponding to the maximum extension ($\gamma =2$) value where the all orders beta function requires the electric theory to start developing an IRFP. This time the condition $X>1$ yields a weaker constraint, i.e. $N_f > 1.5$, with respect to the asymptotic freedom constraint on the lowest value for $N_f$. The trend is different with respect to the case in which we considered magnetic quarks transforming according to the fundamental representation of the $SU(X)$ gauge group.

\subsubsection{$\alpha =4 $, $\beta = 11$ and $\delta =2$}
In this case the solution is:
 \begin{eqnarray}
 X &= & 20N_f- 31 \ , \quad \ell_{A}=0 \ ,  \quad \ell_{D_A} = k_1  = -\ell_{B_A}  \ , \nonumber \\
 && \nonumber \\ 
   \ell_{S}&=&  -2 + 2k_1 + 5k_2 \ , \quad  \ell_{D_S} =  25 +  4 k_1 + 9 k_2  = - \ell_{B_S} \ , \nonumber \\ 
   && \nonumber \\
   \ell_C  & = & 0 \ , \qquad 
  y = 3 \frac{a+ bN_f +c N_f^2}{10(20N_f  - 31)(2N_f - 3)}  \ .\nonumber \\
   && \ \ 
 \end{eqnarray}
 with $k_1$ and $k_2$ integer numbers and 
 \begin{equation}
a = 16- 12 k_1 - 30 k_2 \ , \quad b= 85 -  k_1 + 2k_2 \ , \quad c= 27 + 3 k_1 +4 k_2  \ .
\end{equation}
 Asymptotic freedom requires the previous coefficient to be positive which means: 
 \begin{eqnarray}
 1.59 \leq  N_f   \leq 5.36, \quad  {\rm Dual~asymptotic~freedom~condition}\ .
 \end{eqnarray}
 The lower bound is close again to the value of the critical number of flavors corresponding to the maximum extension ($\gamma =2$) value where the all orders beta function requires the electric theory to start developing an IRFP. The condition $X>1$ yields a  constraint, i.e. $N_f > 1.6$ consistent with the lowest value of the asymptotic freedom window. Note that we have arranged $X$ in such a way that for $X\geq 2$ we recover identically the all order beta function bound for $\gamma=2$.  
 
We were able to find a solution for different values of $\alpha$, in particular for $\alpha =3$ the condition $X\geq 2$  is consistent with the bound of the all orders beta function but for $\gamma =1$ and asymptotic freedom requires $2.15 < N_f < 5.28 $.

{} For duals  with {\it magnetic} quarks in the two index symmetric representation we find more difficult to have a reasonable interpretation of the magnetic baryons, i.e. possessing $B[q^X] = 3n$.  Our findings suggest that duals with fermions in the fundamental representation are, actually, privileged. 

\subsubsection{Summary of this section}
We have found solutions matching the predictions coming from the conjectured all orders beta function also in the case of theories with fermions in the two-index symmetric representation of the $SU(3)$ gauge group. Moreover if one uses dual quarks in the fundamental representation the typical size of the allowed conformal window is consistent with the $\gamma =1$ condition. On the other hand, when using dual quarks in the two-index symmetric representation the size of the conformal window compatible with 't Hooft anomaly matching can extend to match the one obtained using $\gamma =2 $ in the all orders beta function. However the latter case is disfavored by the operator matching conditions given that the $U_V(1)$ charge of magnetic baryons is typically not an integer number of ordinary baryons.

\section{Minimal Conformal Theories: SU(N) with Adjoint Weyl Matter} 
We considered till now only a fixed number of colors since the spectrum of possible composite fermions increases when increasing the number of colors. We turn our attention now to another class of two-index theories for which the dependence on the number of colors, spectrum-wise, is trivial. These are theories with a generic number of Weyl fermions transforming according to the adjoint representation of the underlying $SU(N)$ gauge group.  The associated quantum flavor group is simply $ SU(N_f)$.
We indicate with
$\lambda_{\alpha;a}^i$ the two component left spinor where $\alpha=1,2$
is the spin index, $a=1,...,N^2 -1 $ is the color index while
$i=1,...,N_f$ represents the flavor. We summarize the
transformation properties in the following table:
\begin{table}[h]
\[ \begin{array}{|c| c | c |} \hline
{\rm Fields} &  \left[ SU(N) \right] & SU(N_f)  \\ \hline \hline
\lambda &{\rm Adj} &{\Yfund }  \\
G_{\mu}&{\rm Adj}   &1\\
 \hline \end{array} 
\]
\caption{Field content of an SU(N) gauge theory with quantum global symmetry $SU(N_f)$. }
\end{table}

The  global anomalies are associated to the triangle diagrams featuring at the vertices three $SU(N_f)$ generators. We indicate these anomalies for short with:
\begin{equation}
SU(N_f)^3   \ .
\end{equation}
For a vector like theory there are no further global anomalies. The
cubic anomaly factor, for fermions in the fundamental representation,
is one leading to 
\begin{equation}
SU(N_f)^3 \propto N^2 - 1 \ .
\end{equation}

 We seek solutions of the anomaly matching conditions for a possible dual gauge theory $SO(X)$  featuring 
{\it magnetic} Weyl quarks ${q}$ transforming according to the vector representation of the gauge group. The global symmetry group is then $SU(N_f)$. 
We also add gauge singlet fields built out of the {\it electric} quarks $\lambda$.  The dual spectrum is summarized in table \ref{dualgeneric}.
\begin{table}[h]
\[ \begin{array}{|c| c|c |c|} \hline
{\rm Fields} &\left[ SO(X) \right] & SU(N_f) & \# ~{\rm  of~copies} \\ \hline 
\hline 
 q &\Yfund &{\Yfund } &1 \\
\Lambda &1&{\Yfund}& \ell_\Lambda \\
 \hline \end{array} 
\]
\caption{Massless spectrum of {\it magnetic} quarks and baryons and their  transformation properties under the global symmetry group. The last column represents the multiplicity of each state and each state is a  Weyl fermion.}
\label{dualgeneric}
\end{table}
The gauge singlet state $\Lambda$ is nothing but the gauge singlet built out of the gauge field strength and ${\lambda}$. We can have several copies of $\Lambda$. 

Having defined the possible massless matter content of the gauge theory dual we compute the relevant anomaly:   
 \begin{eqnarray}
SU(N_f)^3 &\propto &  X +  \ell_{\Lambda}= N^2 - 1  \ .  \end{eqnarray}
  The right-hand side is the corresponding value of the anomaly for the electric theory. {}For any $X$ we have a solution which is: 
  \begin{equation} 
  \ell_{\Lambda} =  N^2-1 -X \ . 
  \end{equation}

%

 The one-loop coefficient of the beta function is: 
 \begin{eqnarray}
  \beta_0 = \frac{11}{3}(X-2)   - \frac{2}{3}N_f \ . 
 \end{eqnarray}
 We find that for $X= N_f -1$ asymptotical freedom is lost for: 
 \begin{eqnarray}
  N_{f} \geq \frac{11}{3}  \ , \quad  {\rm Dual~asymptotic~freedom~ and~}N_f~{\rm Weyl~fermions} \ ,
 \end{eqnarray}
 in total agreement with the lower bound of the conformal window obtained by imposing $\gamma=1$ in the all orders beta function. 
 In fact $N_f$ must be larger or equal than four for the dual $SO(N_f-1)$ theory to be a non-abelian gauge theory. Since $N_f$ counts the number of Weyl fermions we have found that the number of Dirac flavors above which we expect any $SU(N)$ gauge theory to develop an infrared fixed point must be equal or larger than two. This is an extremely interesting result since it agrees with earlier analytical expectations obtained using several different analytic methods as well as recent first principle lattice results \cite{Catterall:2007yx,Catterall:2008qk,Hietanen:2008mr,DelDebbio:2008tv,Hietanen:2009az,DelDebbio:2009fd}.

 We have also explored the possibility to introduce dual fermions in the adjoint representation of a SU dual gauge group. Although solutions to the anomaly conditions are straightforward we find that the solution above is the one which better fits the numerical and analytical results. 
 
\section{Conclusion}
We provided the first investigation of the conformal window of nonsuperymmetric gauge theories with sole fermionic matter transforming according to higher dimensional representation of the underlying gauge group. We argued that, if the duals exist, they are gauge theories with fermions transforming according to the defining representation of the dual gauge group. The resulting conformal windows match the one stemming from the all-orders beta function results when taking the anomalous dimension of the fermion mass to be unity. 
 In particular our results for the adjoint representation indicate that for two Dirac flavors any $SU(N)$ gauge theory should enter the conformal window. These results are in excellent agreement with numerical and previous analytical results \cite{Sannino:2004qp,Dietrich:2006cm,Ryttov:2007cx,Sannino:2009aw,Poppitz:2009uq}.  The mapping of higher dimensional representations into duals with fermions in the fundamental representation can be the source of the observed universality of the {\it size} of the various phase diagrams for different representations noted in \cite{Ryttov:2007sr}. 

\subsection*{Aknowledgments}
I gratefully thank  S.~Catterall, M.~T. Frandsen, J.~Giedt,  M.O. J\"arvinen,  T.~A. Ryttov, J.~Schechter and K.~Tuominen  for comments on the manuscript or relevant discussions.

\end{document}